\documentclass[10pt,preprint2]{aastex}

\usepackage{color}

\newcommand{\nuc}[1]{${}^{#1}$}
\newcommand{\nucm}[2]{{}^{#1}{\rm #2}}
\newcommand{\msun}{\ensuremath{M_\odot}}

\newcommand{\tal}{3$\alpha$ reaction}

\newcommand{\pow}[2]{\ensuremath{{#1} \times 10^{#2}}}

\newcommand{\sumin}{\sum_{i=1}^{n}}
\newcommand{\sumjn}{\sum_{j=1}^{n}}
\newcommand{\summ}{\displaystyle \sum_{m=0}^{3}}
\newcommand{\ave}[1]{\left\langle {#1} \right\rangle}
\newcommand{\gumi}{\ensuremath{\Gamma_{\! i}}}
\newcommand{\gamim}{\gumi^{- \frac{m}{3}}}
\newcommand{\thomfer}{\frac{k_{TF}^{2}}{4 k_{F}^{2}}}
\newcommand{\breq}[1]{\ensuremath{Q_{\textrm{#1}}}}
\newcommand{\bref}[1]{\ensuremath{F_{\textrm{#1}}}}
\newcommand{\breg}[1]{\ensuremath{G_{\textrm{#1}}}}
\newcommand{\mui}{\sum \frac{X_{i} Z_{i}}{A_{i}} \rho}
\newcommand{\muii}{\sum_{i} \frac{X_{i} Z_{i}}{A_{i}} \rho}
\newcommand{\popthree}{Pop.III}
\newcommand{\ups}[1]{\ensuremath^{{\textrm{#1}}}}
\newcommand{\lows}[1]{\ensuremath_{{\textrm{#1}}}}
\newcommand{\teff}{\ensuremath{T_{\textrm{eff}}}}
\newcommand{\lhe}{\ensuremath{\log L_{\textrm{He}}}}

\slugcomment{Revised version of 2007/5/22, 4th version}

\shorttitle{Evolution of Low-Mass Population III Stars}
\shortauthors{Suda, Fujimoto and Itoh}

\begin{document}

\title{Evolution of Low-Mass Population III Stars}

\author{Takuma Suda\altaffilmark{1}, Masayuki Y. Fujimoto}
\affil{Department of Physics, Hokkaido University, Sapporo 060-0810, Japan} 
\email{suda@astro1.sci.hokudai.ac.jp}
\and
\author{Naoki Itoh}
\affil{Departments of Physics, Sophia University, Kioi-cho, Chiyoda-ku, Tokoyo 102-8554, Japan}
\email{ n\_itoh@sophia.ac.jp }

\altaffiltext{1}{Meme Media Laboratory, Hokkaido University}


\begin{abstract}
We present the evolutionary models of metal-free stars in the mass range from 0.8 to $1.2 \msun$ with up-to-date input physics.  
The evolution is followed to the onset of hydrogen mixing into a convection, driven by the helium flash at red giant or asymptotic giant branch phase. 
   The models of mass $M \ge 0.9 \msun$ undergo the central hydrogen flash, triggered by the carbon production due to the \tal.  
   We find that the border of the off-center and central ignition of helium core flash falls between 1.1 and 1.2 $\msun$;  the models of mass $M \leq 1.1 \msun$ experience the hydrogen mixing at the tip of red giant branch while the models of $M = 1.2 \msun$ during the helium shell flashes on the asymptotic giant branch. 
   The equation of state for the Coulomb liquid region, where electron conduction and radiation compete, is shown to be important since it affects the thermal state in the helium core and influences the red giant branch evolution. 
   It is also found that the non-resonant term of \tal\ plays an important role, although it has negligible effect in the evolution of stars of younger populations. 
   We compare our models with the computations by several other sets of authors, to confirm the good agreement except for one study which finds the helium ignition much closer to the center with consequences important for subsequent evolution. 
\end{abstract}


\section{Introduction}
Although the initial mass function of metal-free stars formed out of primordial matter has not yet been determined, there is evidence that metal-free stars of low and intermediate mass were formed. 
   For example, the fact that the frequency of stars with strong carbon enhancements is much larger than in the case of population I and II stars \citep[see also the review by Beers \& Christlieb 2005] {ros98,luc05} is consistent with theoretical predictions. 
   One of the most prominent characteristics of models of low-mass and intermediate-mass extremely metal-poor stars (EMPSs) is that they become carbon stars at an earlier stage of evolution than the stars of younger populations \citep[hereinafter FII00]{fuj00}.  
  Recently, it is argued that EMP stars survived to date were formed as a secondary component in the binary systems of massive companions \citep{Komiya07}. 
   As another example, the peculiar abundance characteristics of the most metal poor stars yet discovered \citep{chr02} can be understood in terms of the evolutionary properties of $Z = 0$ or extremely iron-poor models in an intermediate-mass interacting binary \citep{sud04}. 
   Thus, the evolutionary characteristics of $Z = 0$ models have direct relevance to discussions of star formation in the early universe.

EMP stars may alter their initial surface abundances by bringing to the surface products of internal nuclear transformations which occur in unique ways.  
   For initial CNO abundances $\textrm{Z}_{\rm CNO} \lesssim 10^{-7}$, the outer edge of the convective zone, generated by the first off-center helium flash in the hydrogen-exhausted core extends into the hydrogen-rich layers, eventually leading to the enrichment of carbon and nitrogen in the surface (Fujimoto et al.~1990, hereinafter FIH90; Hollowell et al.~1990). 
   This ``He-flash driven deep mixing (He-FDDM)'' mechanism is distinguished from the third dredge-up in asymptotic giant branch (AGB) stars of Populations I and II which enriches surface material in $^{12}$C \citep{ibe75}.  
   Even if we take into account the possible surface pollution by the accretion of metal-rich gas after birth, the interior evolution of metal-poor stellar models is not altered as long as the original CNO abundance satisfies $\textrm{Z}_{\rm CNO} \leq 10^{-8}$ \citep{fuj95}.  

Many calculations of Population III (hereinafter \popthree) star evolution have been published \citep[see FIH90; ][and references therein]{sud04}.
Based on a top heavy initial mass function for the primordial cloud \citet{eze71} first computed sets of massive metal-free models. 
Computations of the evolution of low mass metal-deficient stars were first carried out by \cite{wag74} in order to provide a set of stellar lifetimes for use in calculations of galactic chemical evolution.
Computations of low-mass, zero-metallicity star evolution were performed by \citet{cas75} up to the exhaustion of hydrogen at the center.
\cite{dan82} pursued the evolution of metal-free models of mass 1 $\msun$ to the red giant branch (RGB) phase, but because of a large time step in her computation, did not find any abnormal behavior. 
\cite{gue83} suggested that a convective instability occurs during the central hydrogen-burning phase and FIH90 found this to be the case. 
The ``peculiar'' evolution of $Z=0$ model stars and its interpretation is revealed by the calculations of FIH90 and \cite{hol90}, who followed the evolution of a $Z=0$ star of mass $1 \msun$ and gave explanations for the core He-H flash, a shell He-H flash, and the He-FDDM phenomenon.
By computing in detail the progress of the hydrogen-mixing event and the subsequent evolution, \citet{hol90} first demonstrated that metal-free stars become CN rich carbon stars at the red giant stage.
\citet{cas93} and \citet{cas96} found the core He-H flash in models of $Z=10^{-10}$, but, since their computations did not continue beyond the initiation of the helium core flash, they found no evidence for the additional events just described. 
\cite{wei00} (hereinafter W00) also made computations of low mass metal-free stars, but terminated their computations just after the ignition of the helium core flash while the flash-driven convective zone was still growing in mass.

Computations by \citet[hereinafter S01]{sch01} support the reality
of the He-FDDM phenomenon for low-mass zero metallicity stars, whereas computations by \cite{sie02} (hereinafter SLL02) do not.
While they found the mixing of hydrogen into the helium-flash driven convective zone, the failure of SLL02 to reproduce the He-FDDM phenomenon may be ascribed to their assumption of instantaneous mixing of elements in the helium flash convective zone; in consequence, nuclear energy released from the mixed hydrogen is distributed throughout the entire convective zone and the entropy of the zone is not built up sufficiently for the edge of the convective zone to reach deeply into the hydrogen profile.
Because mixing and burning of hydrogen occur on a very short timescale in the middle of the helium convective zone, following the He-FDDM event properly requires a treatment of time-dependent mixing \citep{hol90}. 
\citet{hol90} first derived the surface composition caused by the He-FDDM phenomenon for a $Z = 0$ model of mass $M = 1 \msun$. 
\citet{sch02} also derived the surface composition for a model of mass $M = 0.82 \msun$ and compared the results with the observations of EMPSs.

After the discovery of HE0107-5240 \citep{chr02}, which held the distinction of being the most metal-poor giant known before the discovery of HE1327-2326 \citep{fre05}, models of EMPSs were computed in an effort to determine the evolutionary history of the observed star. 
Both \citet[][ hereinafter P04]{pic04} and \citet{wei04} found the He-FDDM phenomenon for $Z=0$ models of mass $M = 0.8 \msun$ and $M = 0.82 \msun$, respectively.

All but one of the works just cited find the core He-H flash; thus far, only Fujimoto and his coworkers have found the shell He-H flash driven by a violent CN-cycle burning at the base of the hydrogen-burning shell. 
Why various groups find different evolutionary characteristics for similar masses and compositions has not yet been fully understood. 
One reason for this is that the computation of $Z=0$ models requires not only a careful treatment of the input physics but also a careful numerical solution.
In particular, the treatment of convective regions requires consideration of many complicated factors including uncertain parameters. 
S01 made comparisons among models by changing parameters related to abundances, diffusion, stellar mass, and the mixing length algorithm for convection, and suggestions have been made for the causes of differences such as radiative and/or conductive opacities and neutrino energy-loss rates (W00; S01; SLL02). 
However, we are not satisfied that the basic causes for differences have been pinpointed satisfactorily.

In this paper, we examine the differences in the evolutionary characteristics of low-mass $Z = 0$ models by taking into account the choice not only of the radiative and conductive opacity and of neutrino energy-loss rates, but also of nuclear reaction rates. 
Thanks to the many efforts until today, useful sets of input physics are available and they have been more and more accurate for numerical computations. 
  But we think it is worth trying to check the dependence on the input physics and to elucidate the structural characteristics in detail because these efforts are still going on and because the evolution of low-mass Popualtion III stars has not been established yet. 
  In particular, as for the three alpha reaction rates, the determination of resonance states in \nuc{12}C is still controversial \citep{fyn05}; 
   the simulations of three body interactions are challenging, and yet, become feasible task in near future thanks to the improved computer resources \citep[see e.g., ][and the references there]{kur05}. 

In the next section, we elaborate the input physics adopted in the stellar evolution program. 
In \S 3, we present the result of computations of evolution of low mass $Z=0$ model stars with updated input physics, and discuss the model characteristics and their dependences on the input physics including the resonant and non-resonant reaction rates of 3 $\alpha$ reactions. 
  In section 4, comparisons with other works with the different input physics taken into account, and, in \S 5, we summarize conclusions.

\section{The Computational Program}

The original program to compute stellar evolution was constructed by \citet{ibe65} and has been modified periodically \citep[e.g., see][]{ibe75,ibe92}. 
The size of each mass shell and the time step are regulated, respectively, by the gradients with respect to space and time of structure and composition variables.
Typically, 200-300 mesh points are required for main sequence models and 700-800 for core helium-burning models with hydrogen-burning shells.
The number of equilibrium models required to follow evolution from the zero-age main sequence to the end of RGB or AGB phase varies from 5000 and 30000, with the exact number depending on initial mass. 
The stellar structure equations are typically satisfied to better than one part in $10^{5}$.

In the program, the radiative opacity is obtained by interpolation in OPAL tables \citep{igl96} and in tables by \citet[][ hereinafter, AF94 tables]{ale94} and the conductive opacity is from \citet{ito83}. 
The equation of state involves fits by \cite{ibe92} to work by \cite{abe59}, \cite{bow60}, \cite{sla80}, \cite{sla82}, \cite{iye82}, \cite{han73}, \cite{han78}, \cite{coh55}, and \cite{car61}.
Neutrino energy-loss rates are from \citet[][ in the following I96]{ito96} for plasma, photo, pair and bremsstrahlung processes.
In order to compare with other works, we make use of two sets of nuclear reaction rates: those given by \citet[][ in the following CF88]{cau88} and those given in the latest NACRE compilation \citep{ang99}.
Nuclear screening factors for weak and strong screening are also taken into consideration using standard prescriptions \citep[see, e.g.,][]{bohm58}, but only weak screening is dominant in the actual computational range in this work. 
Nine nuclear species are considered: \nuc{1}H, \nuc{3}He, \nuc{4}He, \nuc{12}C, \nuc{14}N, \nuc{16}O, \nuc{18}O, \nuc{22}Ne, and \nuc{25}Mg. 
Abundances of these isotopes are determined by 16 nuclear reactions which include proton, electron, and alpha captures.

In regions of high temperature and high density which are not covered by OPAL tables, we use the analytical radiative opacities from the fits by \cite{ibe75} to \cite{cox70a,cox70b} ones.
At table boundaries, no interpolation is made between table values and analytical values. 
Hence, jumps in the radiative opacity occur at table boundaries, but jumps are normally smaller than a factor of 2 and do not seriously affect model convergence, primarily because, in regions not covered by the tables, the overall opacity is dominated by electron conductivity. 
At low temperature and low density, interpolation between OPAL and AF94 tables is accomplished by setting
$\kappa=\kappa_{\rm OPAL}\ (1-\Theta)+\kappa_{\rm AF}\ \Theta$, where
$\Theta=\sin^2{(\pi/2)(T-T_1)/(T_2-T_1)}$ for the temperature $T$
between $T_1=8000$ K and $T_2=10000$ K, and $\kappa_{\rm OPAL}$ and $\kappa_{\rm AF}$ are OPAL and AF94 opacities, respectively.
In surface layers of all models discussed here, AF94 tables provide opacities for all densities and temperatures encountered. 
In regions of overlap, OPAL and AF94 opacities agree rather well, so switching from one table to the other introduces little uncertainty.

\begin{figure}
\epsscale{.80}
\plotone{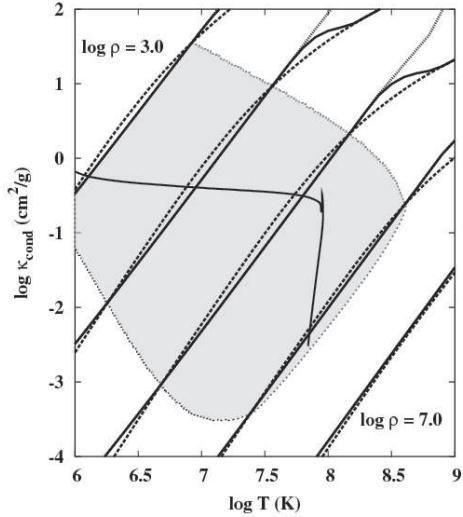}
\caption{
Interpolation of conductive opacities between I83 and \citet{hub69}as approximated by \citet{ibe75} for pure $\nucm{4}{He}$ composition.
Conductive opacities by I83 (dotted lines) and by Iben's fitting formulae (dashed lines) are interpolated (solid lines) in the range $0.5 \leq T/T_{F} \leq 2.0$, where $T_{F}$ is the Fermi temperature.
Each Line gives the conductive opacity for a constant density overthe range 10$^{3}$ - 10$^{7}$ g cm$^{-3}$. 
   The shaded area surrounded by dotted line represents where the conductivity by I83 becomes effective for density greater than given by Iben's fitting formulae. 
  Thin solid line shows the distribution of opacities in the 10nac model during the helium core flash as a function of temperature.
}
\label{fig:cop}
\end{figure}

Although the quantum correction to the conductivity at low temperatures has been calculated \citep{mit84}, it is now believed that the approximation employed is not appropriate at the temperatures considered. 
Therefore, we use the analytic fits devised by I83 for the conductive opacity at low temperature; 
the liquid metal phase for various elemental compositions as well as strong degeneracy are taken into consideration.
Strong degeneracy prevails if $T \ll T_{\rm F}$, where $T_{\rm F}(K) = 5.930 \times 10^{9}\
((1+1.018(\rho_{6}/\mu_e)^{2/3} )^{1/2} -1 )$, $\rho_{6}$ is the
density in units of $10^{6}$ g cm$^{-3}$, and $\mu_e$ is the electron molecular weight. For the actual computations, since their approximations are good for $T \lesssim T_{\rm F}$, we adopt the analytical approximations of \citet{ito83} for $T \leq 0.5~T_{\rm F}$.
For $T \geq 2~T_{\rm F}$, we adopt the conductive opacity used in the fits by \citet{ibe75} to calculations by \cite{hub69} for non-relativistic electrons and to \cite{can70} for relativistic electrons.
For intermediate temperatures, interpolation between the two approximations is achieved by ``sine square'' weighting, analogous to that used in interpolating between OPAL and AF tabular opacities.
Finally, the conductive opacity is subjected to a limit described in the Appendix.
Interpolated conductive opacities are given in Figure~\ref{fig:cop}, which covers the overall range of application for stellar evolution.

For neutrino energy-loss rates, we adopt the fitting formulae presented by I96, who include pair, photo, plasma, and bremsstrahlung with weak degeneracy, the liquid-metal phase with low-temperature quantum corrections and the crystalline lattice phase, and the recombination neutrino processes. 
Among these latter processes, we do not adopt the low-temperature correction to the liquid-metal phase as calculated by \cite{ito84} since the corrections are generally small, as stated in I96.
For $T \leq 10^{7}$ K, we assume that neutrino energy-losses of all kinds can be neglected. 
The treatment of a multicomponent gas, which is important for bremsstrahlung, is considered in applying the neutrino energy-loss rates of I96, as described in the Appendix.

Mass loss is neglected in our work because we are interested primarily in the stellar interior and the evolution of the helium core is not affected by modest surface mass loss. 
To determine the temperature gradient in convective regions, we use the standard mixing length recipe by \cite{bohm58} \citep[see, e.g., ][]{cox68} and the mixing length is taken to be 1.5 times the local pressure scale height.
Neither overshooting nor semi-convection are considered.

In our calculations we focus on the evolutionary trajectory up to the
He-FDDM event and on the thermal structure of a star at stages of interest along its path to this event; 
our computations are terminated at the onset of hydrogen mixing into the convective zone driven by a helium flash.
This event can occur during an off-center helium flash, 
and/or at the beginning of the thermally pulsing AGB (TPAGB) phase, depending on the initial mass and input physics.
Subsequent evolution after the mixing event is beyond the scope of thispaper and will be discussed in detail in a separate paper (Suda, Fujimoto, \& Iben, in preparation).

\section{Evolution of Low-Mass \popthree\ Stars}

The evolution of $Z = 0$ stars has been computed from the zero-age main sequence (ZAMS) through the RGB and/or through the TPAGB phase for model masses of $0.8 - {1.2} \msun$.
     For all models, the initial chemical composition is $X = 0.767$, $Y = 0.233$, and $Z=0$, and the initial abundance by mass of \nuc{3}He is $2 \times 10^{-5}$. 
These abundances are based on models of big bang nucleosynthesis and are the same as those chosen by FII00. 

Table~\ref{tab:model} lists the models computed with all the input physics updated for this paper. 
  The first two columns give, respectively, the model identifier and the initial model mass;  the labels ``nac'' and ``cf'' mean that the model has been computed with the choice of nuclear reaction rates of NACRE and CF88, respectively.
   The third to the ninth columns give the effective temperature and surface luminosity at the turn-off point and at the tip of the RGB, the times to reach these two stages, respectively.  

\begin{figure}
\plotone{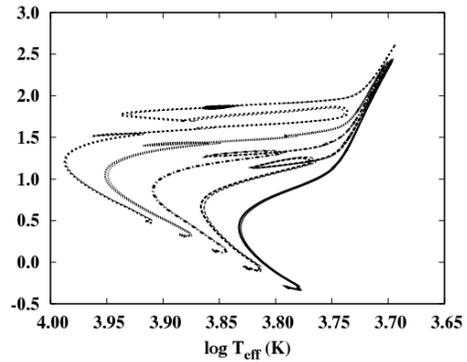}
\caption{
H-R diagram for models of masses in the range from $0.8$ to $1.2 \msun$ in steps of $0.1 \msun$ with the nuclear reaction rates given by NACRE[][; thick lines]{aug99} and by \citet[][; thin lines]{cau88}. 
   Evolutionary tracks are plotted from the zero-age main-sequence to the hydrogen-mixing event at the tip of RGB (for $0.8 \msun \ge M \ge 1.1 \msun$) and to the early thermal-pulsating AGB (for $M=1.2 \msun$), respectively. 
}
\label{fig:hrd}
\end{figure}

Figures~\ref{fig:hrd} and \ref{fig:rhot} show, respectively, evolutionary tracks in the H-R diagram and in the central-density and central-temperature plane for ''cf'' and ''nac'' models of mass in the range $0.8 -1.2 \msun$. 
   All models are terminated at the onset of hydrogen mixing episodes.
   We found that $M \leq 1.1 \msun$ models ignite the helium burning in the off-center shell and the engulfment of hydrogen by the flash-driven convection occurs just after the peak of the major core helium-flash at the tip of RGB. 
   On the other hand, $1.2 \msun$ models undergo the helium core flash at the center without the hydrogen mixing, and encounter the hydrogen mixing in the helium-flash convection during the helium shell flash at the beginning of thermally pulsating AGB (TP-AGB) phase. 
   The two models display almost the same trajectory irrespective of the choice of nuclear reactions, with the small differences stemming from the difference in the \tal\ rates, less than a factor of 2 in the temperature range $\log T = 7.8 - 7.9$.  

The evolutionary characteristics of the core helium flash and the hydrogen mixing event are summarized in Table~\ref{tab:he-flash}. 
   Each column gives the model identifier defined in Table~\ref{tab:model}, the helium core mass $M_{1}\ups{max}$ and the helium burning rate $L\lows{He}\ups{max}$ when the helium burning rate reaches maximum, the mass coordinate $M\lows{BCS}$ and the maximum temperature  $T\lows{BCS}\ups{max}$ at the base of the convective shell, driven by the core helium flash, the helium burning rate $L\lows{He}\ups{mix}$ at the onset of hydrogen mixing at the RGB, and the time intervals, $\Delta t^\prime$ and $\Delta t\lows{mix}$, to it from the appearance of helium flash-driven convection and from the stage of maximum helium burning, respectively. 
   The mass $M_1$ of helium core is defined as the mass coordinate where the abundance of hydrogen is half of the surface abundance of hydrogen.

In this section, we discuss in detail the evolutionary behavior of \popthree\ models with respect to the differences in initial mass and input physics.

\subsection{Hydrogen Burning Phase}

Variations in several quantities characterizing model stars are summarized in Figure~\ref{fig:phys} as a function of the hydrogen abundance at the center.  
   In a \popthree\ model, due to the absence of CNO catalysts, p-p chain reactions are initially the only mode of energy generation by hydrogen burning. 
   Because of the weak temperature dependence of the energy-generation rate, the central temperature keeps rising as the hydrogen abundance decreases (top panel of Fig. \ref{fig:phys}). 
   As the central temperature increases, the \tal\ gradually becomes active and CNO-cycle reactions begin to occur.  
   Eventually, CNO-cycle reactions dominate the p-p chains with regard to total energy production. 
   When this first occurs, the central abundance by mass $X \lows{CNO}$ of catalysts is $10^{-11} - 10^{-9}$ at $\log T_{c} \simeq 7.8 - 7.9$ (top and middle panels of Fig. \ref{fig:phys}). 
   Because of lower central temperatures, the less massive the star, the later is the evolutionary stage (and the smaller is the central hydrogen abundance) at which the transition from burning dominated by the p-p chains to burning dominated by the CNO-cycle reactions takes place. 
   For $M \leq 0.8 \msun$, hydrogen is depleted in the center before a transition can take place. 
   After the CNO cycle takes over as the main source of energy generation, because of the strong temperature dependence of CNO-cycle reactions, the central temperature remains nearly constant. 
   At the same time, the core expands because of the central concentration of energy generation and the rate of production of catalysts slows down; 
   the abundances of catalysts saturate at $X \lows{CNO} \simeq 10^{-10} - 10^{-8}$.

The transition to the CNO-cycle dominated phase is accompanied by the formation of a convective zone which develops outward from the center as described in the lower panel of Fig.~\ref{fig:phys}. 
   The growth of the convective core is also evident in the middle panel of Fig.~\ref{fig:phys}, which shows that the hydrogen abundance at the center, $X_{c}$, stops decreasing monotonically and increases for a time as evolution progresses. 
   Central convection is caused by a thermonuclear runaway and persists after the transition; due to the larger temperature dependence of the CNO-cycle energy-generation rate, energy generation is highly concentrated toward the center. 
   Both the inward mixing of hydrogen from outer hydrogen-rich layers and the outward mixing of CNO elements generated near the center during the thermonuclear runaway amplify the average hydrogen-burning rate over the region encompassed by convection relative to the average rate in the absence of convection.

\begin{figure}
\plotone{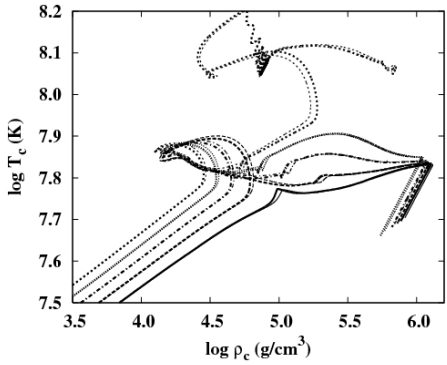}
\caption{
Evolutionary tracks of model centers in the temperature-density plane. 
The models are the same as in Fig.\ref{fig:hrd}.
}
\label{fig:rhot}
\end{figure}

\begin{figure}
\plotone{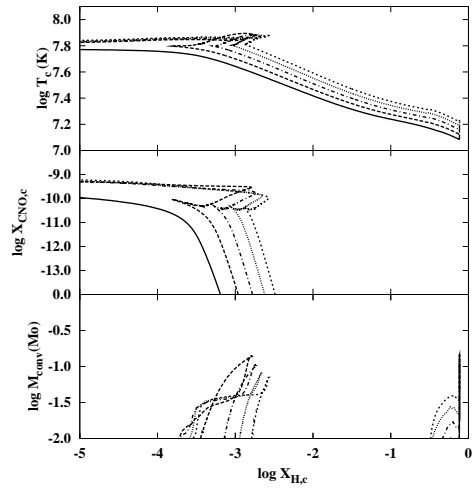}
\caption{
Variations in several physical quantities as a function of the hydrogen abundance at the center. 
   In order from top to bottom, panels show the central temperature, mass fraction of the sum of CNO elements at the center, and the mass of the convective region. 
   Each line represents the time variation of 0.8 (solid lines), 0.9 (dashed lines), 1.0 (dot-dashed lines), 1.1 (dotted lines), and 1.2 $\msun$ (short-dashed lines) model, respectively.
}
\label{fig:phys}
\end{figure}

For stars of mass $0.9 \leq M / \msun \leq 1.2$, the transition from the p-p chain dominated phase to the CNO-cycle dominated phase is delayed until the core has already begun to contract rapidly and electrons have begun to become degenerate at the center;  
   the maximum energy generation has already shifted away from the center and core contraction has initiated the expansion of the envelope. 
   A thermonuclear runaway called the core helium-hydrogen (He-H) flash takes place \citep{fuj90}. 
   In Fig.~\ref{fig:rhot}, a first increase in the central temperature with decreasing central density indicates that the electron degeneracy is lifted; 
   then, the temperature turns to decrease with the density so as to settle in the thermal equilibrium state where the nuclear energy generation balances the energy loss from the core.   
   In this runaway, helium burning plays a key role through the production of CNO-cycle catalysts. 
   Because of the strong temperature dependence of CNO-cycle reaction rates, the flash starts even when the contribution of the CNO cycle to the total energy-generation rate is smaller than that of p-p chain reactions in central regions.  
   Smaller mass stars experience stronger flashes since the central entropy is smaller and the electron degeneracy is stronger at the onset of the thermonuclear runaway. 
   Thus, the temperature reaches larger, leading to a greater production of CNO catalysts and a greater extension of convection. 
   For stars of mass $M \leq 0.8 \msun$, the central temperature does not become large enough for the production of sufficient carbon to activate the CNO cycle, irrespective of the choice of reaction rates. 
   We see a small hump along the trajectories of $M = 0.8 \msun$ models in Fig.~\ref{fig:rhot}, which marks where the exhaustion of hydrogen at the center quenches a thermonuclear runaway.

These evolutionary characteristics during core hydrogen burning are common to all of the models, regardless of the adopted nuclear reaction rates since the evolutionary tracks prior to the transition are dominated by the p-p chain reactions with relatively small temperature dependences.  
   All the models of masses in the range $0.9 \leq M / \msun \leq 1.2$ experience the core He-H flash.  
   We may notice that the differences in the ignition temperatures are rather small because of the strong temperature dependence of the \tal\ rate, the flash grows weaker for models of larger mass.

Core convection lasts until hydrogen is almost exhausted at the center ($X_{c} \lesssim 10^{-4}$).  
   After the model settles in the thermal equilibrium, the central density increases again as the hydrogen abundance decreases. 
   The central temperature at first rises, and then, decreases as the nuclear energy-generation rate decreases at the center (when $X_{c} < 2 - 3 \times 10^{-4}$). 
   Note that since the nuclear energy generation is dominated by contribution from off-center burning, the model at this phase tends to have an isothermal core with the central temperature directly reflecting the temperature in the H-burning shell. 
   In Fig.~\ref{fig:rhot}, we can see the rise in central temperature with the growth of core, a sudden jump in temperature occurring when the hydrogen burning shell passes the shells occupied by the flash convection, and hence, have larger CNO abundances.  
   
The resultant loop, as seen in Fig.~\ref{fig:rhot}, is characteristics of flash, implying that He-H core flash eventually exerts work through the expansion and contraction of core, during which the envelope first contacts and then expands again, as seen in Fig.~\ref{fig:hrd}. 
   In the subsequent evolution, our models do not encounter the phenomena, so called ``shell He-H flash'' at the base of hydrogen burning phase, as FIH90 discovered. 
   FIH90 discuss the behavior of hydrogen burning shell in the $Z=0$ environment and found the convective instability by the production of carbon and of nuclear energy by CN cycles. 
   This is the counterpart of core He-H flash, i.e., electrons are degenerate at the base of hydrogen burning shell, which causes He-H flash. 
   However, no other groups than Fujimoto and his collaborators find such a event. 
   The different consequence is due to the consideration of non-resonant effect in \tal\ rate as we will discuss in \S \ref{sec:tal}. 

\subsection{Helium Burning Phase}\label{sec:heburn}

Characteristic properties of the helium core flash are given in Table~\ref{tab:he-flash}. 
   The thermal structure of the core when the helium core flash is ignited are determined by (1) neutrino energy losses which produce cooling, (2) the radiative and conductive opacity which controls heat flow, and (3) the temperature and energy generation in the hydrogen-burning shell which affects the heating due to gravitational compression. 
   If neutrino cooling in the central region is effective enough to produce a strong positive temperature gradient, the helium core flash is ignited off center. 
   Model mass must be larger than $M \geq 1.2 \msun$ for central helium ignition to occur (Fig.~\ref{fig:rhot}). 
   The mass $M_{1}$ of the helium core when the helium core flash occurs differs greatly between the off-center and central ignition cases, as seen from the third column of Table~\ref{tab:he-flash}. 
   Central ignition occurs before neutrino cooling becomes appreciable, and, hence, core masses at ignition are small:  $M_{1} \lesssim 0.4 \msun$.  
   Once neutrino cooling becomes effective in the central region, off-center ignition is delayed until the core mass is much larger, namely, $M_1 \gtrsim 0.5 \msun$.

Figure~\ref{fig:str} shows the interior temperature as a function of density for models of mass 0.8 $\msun$ and core mass $M_{1} = 0.49 \msun$. 
   A constant temperature ``plateau'' spreads inward from the base of the hydrogen-burning shell. 
   This implies that the thermal structure of the helium core results mainly from the temperature in the hydrogen burning shell and neutrino energy losses in central regions \citep{fuj84}. 
   Along with a decrease in the initial abundance of CNO elements, the hydrogen-burning rate decreases, and, in order to compensate for this, the temperature in the hydrogen-burning shell increases, which makes the contribution of compressional heating smaller. 
   Accordingly, an isothermal plateau develops in the region where the radiative heat transport dominates over electron conduction for the metallicity of $[{\rm Fe}/{\rm H}] \lesssim -5$ \citep[see ][]{fuj95}. 
   In our models, the compressional heating plays a small part to raise the maximum temperature in the helium zone slightly higher than the temperature in the hydrogen burning shell, as seen from the lowest mass model;  
   note that for the massive models, the helium burning already contributes appreciably to increase the temperatures in the right shoulder of structure lines in this figure.  
   This behavior contrasts with that of model stars of younger population in which the compressional heating play the dominant part in determining the maximum temperature in the helium core and makes it much larger than the temperature in the hydrogen-burning shell.  
   Because of high temperature in the hydrogen burning shell, therefore, $Z=0$ model stars experience central ignition at smaller initial masses than do model stars of younger populations for which the minimum initial mass for central helium ignition is $\sim 2.5 \msun$. 
   While the density and temperature of the hydrogen burning shell are in local maximum just before the helium ignition, hot CNO-cycle is not still effective.
   At this stage, the nuclear timescale of $\nucm{13}{N}$ against proton capture reaction ($\sim 10000$ sec) is much larger than that against $\beta$-decay reaction (863 sec) and is negligible in the outcome of neither the nucleosynthesis nor the nuclear energy output.

If neutrino energy losses are sufficiently effective before helium is ignited, the central region cools, and helium is ignited off-center as is the case in low-mass stars of younger populations. 
   In the central region of the model, plasma neutrinos are the dominant neutrino energy-loss mechanism and produce a steep gradient in the rate of released energy. 
   On the other hand, the conduction, which is important in the electron-degenerate core, transport the energy towards the center where the neutrino loss works. 
   Consequently, both conductivity and neutrino energy-loss rates promote cooling of the core and, thus, delay the off-center ignition of helium until a larger core mass gives rise to a larger hydrogen-shell burning rate and a larger temperature in the hydrogen-burning shell, as seen from Table~\ref{tab:he-flash}.   

In all of the models which experience off-center ignition, convection driven by helium burning extends into the upper hydrogen-rich layers during the decay phase of the core helium flash, because of smaller entropy in the hydrogen burning shell, as shown by \cite{fuj90,fuj95}. 
   The ingestion of hydrogen into the helium convective zone begins a sequence of events that leads to the enrichment of the surface with carbon and nitrogen \citep{hol90,fuj00,sch02,pic04,wei04}. 
   Characteristics of hydrogen mixing are also given in Table~\ref{tab:he-flash}. 
   The ingestion of hydrogen occurs within a matter of days after the helium-burning luminosity reaches its peak.
   In the models undergoing central helium burning, we do not find a hydrogen-mixing event;  
   The core helium flash is rather weak (see $L_{\rm He}^{\rm max}$ in Table~\ref{tab:he-flash}), which makes it difficult for the outer edge of the convective region driven by helium burning to reach the hydrogen-containing layer \citep{fuj77}. 
   For the models of $M=1.2 \msun$, we follow the evolution through the thermally pulsing AGB phase to find that the He-FDDM is triggered during the helium shell flashes. 

\section{Discussion and Comparison with Other Works}

In this section, we compare the models in the literature with our models adopting similar input physics to see which cause the dominant effect on the differences in the evolution and to confirm the correctness of numerical computations.
   In addition, we compare the models without non-resonant effect for $\alpha$-capture reactions to see the influence of uncertainty in \tal\ rates on the stellar structure at $Z=0$. 

\subsection{Comparison with $0.8 \msun$ Models}

A model comparable with our $0.8 \msun$ model is Model~1 of \citet[][P04]{pic04}, of composition $Y = 0.23$ and $Z = 0$.
In their computations, release 4.98 of FRANEC was used and time-dependent convective mixing was calculated; neutrino energy-loss rates by plasma-neutrino emission were modified, with consequences being reported as minimal.
Reaction rates and conductive opacities are, respectively, common with our models.
The neutrino energy-loss rates are common with our model for photo- and pair-neutrino processes, but they use energy-loss rate of \citet{dic76} for bremsstrahlung and of \citet{bea67} for recombination processes.  
   For plasma neutrino energy losses, they adopt an energy-loss rate \citep{esp03} which differs only slightly from I96 in the temperature and density ranges relevant to the ignition of the helium flash.

Although there are some differences in adopted input physics, the evolution of the helium core flash of the P04 model is similar to that of our 08cf model. 
     The helium core mass at the onset of the core helium flash is the same in both models, namely, $M_{1} = 0.52 \msun$. 
When CNO-cycle reactions are the dominant contributors to the hydrogen-burning luminosity, so that the hydrogen profile is very steep, the quantity $M_{\rm{He}}$ defined by P04 as the mass of the helium core when the maximum hydrogen-burning luminosity is reached is nearly the same as the quantity $M_{1}$ we have defined as the mass of the helium core when the core helium flash begins.  
The maximum helium-burning luminosities differ by less than a factor of two, being $L_{\rm He} = \pow{1.2}{10} L_{\sun}$ in the P04 model and $L_{\rm He} = \pow{7.6}{9} L_{\sun}$ in our model 08cf. 
The mass at the outer edge of the convective shell at the onset of hydrogen mixing is the same in both cases, namely, $0.506 \msun$.
The values of $\Delta t_{\rm{mix}}$ (see Table \ref{tab:he-flash}) and $X_{\rm C}$, the mass fraction of carbon in the helium convective shell, are also comparable: 
$\Delta t_{\rm{mix}} = \pow{2.1}{5}$ sec in the P04 model versus $\pow{1.87}{5}$ sec in model 08cf, and $X_{\rm C} = \pow{4.15}{-2}$ in the P04 model versus $X_{\rm C} = \pow{4.28}{-2}$ in model 08cf.
There is a large difference in model characteristics when the helium convective shell first appears;  
in the P04 model, the convective shell driven by helium burning appears at a mass shell $M_{\rm BCS} = 0.348 \msun$ when $L_{\rm{He}} = 0.658 L_{\sun}$, while, in our model 08cf, $M_{\rm BCS} = 0.3825 \msun$ when $L_{\rm He} = \pow{1.57}{2} L_{\sun}$.  
    The factor of 200 difference in the helium-burning luminosity when shell convection begins is probably simply a typographical error in P04, an interpretation reinforced by the fact that the time for the helium-burning luminosity to reach its maximum value is almost the same in the P04 model (723 yr) and in ours (715 yr).
It takes more than $\sim 3\times 10^4$ yr for the helium-burning rate to increase by a factor of 200 in this range. 

The model of $0.8 \msun$ and $Z=0$ in FII00 is based on those of \citet{fuj95} (hereafter F95) and the comparable results are given in their Table 1. 
   In F95, the values of  $M_{1} \ups{max}$, $M_{\rm BCS}$, and $\lhe \ups{max} (L_\odot) $ are $0.5116 \msun$, $0.3705 \msun$, and 9.983, respectively.  
   This core mass $M_1$ coincides with our 08nac very closely despite the differences in the input physics;  
   F95 took into account only the resonant \tal\ reactions \citep{aus71}, which is smaller by a factor of $\sim 2.4$ than the NACRE rate at the relevant temperature range ($\log T \simeq 7.94$). 
   The smaller helium burning rate tends to delay the ignition of helium core flash.  
   On the other hand, the I83 formulae adopted here give larger conductivity than the Iben's fitting formulae used by F95 in the region of coulomb-liquid regime where the maximum temperature in the helium zone occurs (see fig.~\ref{fig:cop}), which works to delay the helium ignition due to the enhanced cooling of helium zone through the inward heat conduction in our model.  
   These two effects compensate for each other, while the effect of  larger conduction is manifest in inner ignition (or in smaller  $M_{\rm BCS}$) in the 08nac model.  
   In actuality, the 08cf model with the same input physics as the 08nac model except for the nuclear reaction rates results in a larger core mass than the 08nac model. This is because the cf88 \tal\ rate is smaller by $20 \%$ than the NACRE \tal\ rate around the temperature relevant here.  
   In any case, the dependence of $M_1$ on the nuclear reaction rate is very small because of strong temperature dependence of \tal\ rate. 
   It is also worth noting that the non-resonant reaction \citep{nom85} has little to do with the ignition of helium core flash because of rather high temperatures in the helium zone $\log T > 7.9$, although it takes a major part in the later phase of core hydrogen burning and the omission delays the depletion of hydrogen until higher temperature is reached in the center ($\Delta T_{\rm c} \simeq  0.04$ and $\Delta \rho_{\rm c} \simeq  0.25$).  

Since our calculation does not follow the burning of mixed-in hydrogen, we cannot assess the results of FII00 and P04 with regard to the occurrence of a hydrogen-burning flash in the middle of helium convective zone, the splitting into two convective shells, and the merging of the upper convective shell with the surface convective zone. 
   This remains for further investigation and will be discussed in a separate paper (Suda, Fujimoto, \& Iben, in preparation).

\subsection{Comparison with $1 \msun$ Models}

In this section, we compare our results with those of FIH90, W00, S01, and SLL02 for the evolution to the beginning of the core helium flash of models of mass $1 \msun$ and initial composition $Z=0$. 
   In all of the cited calculations, the core He-H flash occurs, although the size of the blue loop differs among the different works.

\begin{figure}
\plotone{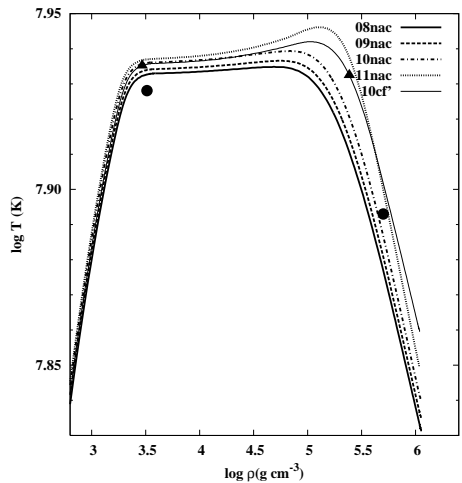}
\caption{
Stellar structure in the temperature-density plane for various models at an evolutionary stage corresponding to a helium core mass of $M_{1} = 0.49 \msun$. 
   Model names are described in Table~\ref{tab:model}. 
   The two filled circles show $\rho$ and $T$ at the hydrogen-burning shell and at a shell of mass $M_{r} = 0.151 \msun$ in the 10cf$^\prime$ model for which $M_{1} = 0.482 \msun$, the same core mass as in S01 at the onset of helium ignition. 
   For comparison, the thin solid line describes the 10cf$^\prime$ model a few Myr before helium ignition. 
   Triangles on the line locate the ignition point (right hand side) and the hydrogen-burning shell (left hand side).
}
\label{fig:str}
\end{figure}

We first compare our results with the model of FIH90.  
   The distinctive feature of FIH90 model is the He-H shell flash during the hydrogen shell-burning, as stated in the introduction. 
   We will show in the following subsection that the instability of the hydrogen shell burning and the resultant shell flash are solely attributable to the exclusion of non-resonant rate of \tal\ reactions \citep{nom85}\footnote{The computation of FIH90 was done at the University of Tokyo Observatory in 1984 when one of authors (I.I., jr.) visited it as JSPS fellow.}.  
   Furthermore, the FIH90 model ended in a significantly larger core mass at the ignition of helium core flash ($M_{1} = 0.528 \msun$). 
   It is even larger as compared with the $0.8 \msun$ model of FII00 with the same input physics except for the equation of state (EOS), despite the general tendency of decrease for larger stellar masses as seen in Table.~{\ref{tab:he-flash};   
  Since these two computations differ only in the equation of state (EOS) for a Coulomb liquid and solid among the input physics \citep[see][for the adopted the EOS]{ibe92}, the main reason for the larger core mass may be the larger radius of the helium core in the FIH90 model (see their Table 1 in FIH90), resulting from the difference in the adopted Coulomb corrections in the EOS.  
    A larger core radius implies a smaller gravity of the core, and hence, a smaller temperature in the hydrogen-burning shell for a given core mass, to defer the ignition of helium core flash until a larger core mass is achieved.   
   In actuality, in FIH90 model, a He-H shell flash is postponed until the core mass grows as large as $M_1 = 0.505 \msun$ and ignited at a low density in the bottom of hydrogen burning shell under a flat configuration ($V \gg 4$) but under non electron-degenerate conditions \citep[see their Fig.~2 in][]{fuj82}. 
   Because of a large core mass, the He-H shell flash has little effect on the thermal state of the inner core and FIH90 find off-center ignition of the helium core flash.  

Next, we compare with the work by \citet[][ W00]{wei00} who use a code which differs from the one used by \citet[][ S01]{sch01} with regard to the EOS \citep{str88}, reaction rates \citep[][which include the non-resonant term in the \tal\ rate]{thi87}, and the radiative opacity \citep[old version of OPAL,][]{rog92,igl92}. 
   The conductive opacity and the neutrino energy-loss rates are not described in their paper. 
   Since they do not follow the progress of the helium core flash, they do not find the He-FDDM event while helium is ignited off center. 
   Their value of $t \lows{TO} = 6.31$ Gyr is close to that of our model 10cf (6.54 Gyr);  
   the small difference may be due to the use of different versions of OPAL opacities. 
   As for the core He-H flash, their values for the location of the outer edge of the central convective zone $M_{ECS} = 0.11 \msun$ and for the helium-burning luminosity $L \lows{He} \ups{max} = 2.57 \times 10^{-7} L_{\sun}$ are similar to our values of $M_{ECS} = 0.115 \msun$ and $L \lows{He} \ups{max} = 3.27 \times 10^{-7} L_{\sun}$ for our model 10cf. 
   Element abundances at maximum nuclear burning luminosity are also comparable; 
   W00 find $X_{12} = \pow{6.50}{-12}$, $X_{14} = \pow{2.19}{-10}$, and $X_{16} = \pow{2.77}{-12}$, and we find $X_{12} = \pow{7.90}{-12}$, $X_{14} = \pow{1.20}{-10}$, and $X_{16} = \pow{8.44}{-13}$. 
   These abundances are influenced slightly by the choice of time step. 
   At the tip of the RGB, we can compare with their ``canonical'' model that element diffusion is not considered in their work;  
   their $M_{1} \ups{tip} = 0.497 \msun$ and $\log (L_s^{\rm tip} /L_{\sun}) = 2.357$ differ slightly from our 10cf model: $M_1^{\rm tip}  = 0.5054 \msun$ and $\log (L_s^{\rm tip}/L_{\sun}) = 2.413$ (See Tables~\ref{tab:model} and \ref{tab:he-flash}). 

In the S01 calculations, evolution is followed from the main sequence to the TPAGB phase. 
   The S01 input physics differs from ours with regard to the EOS in regions of electron degeneracy \citep{kip90}, the energy-loss rates for photo, pair, and plasma neutrinos \citep{mun85}, and weak screening for nuclear reactions (they use the Salpeter formula). 
   For the EOS in core regions, they adopt a simplified equation of state for a degenerate electron gas \citep{kip90} but they provide no
explicit statement as to their treatment of the ion gas.
   Their nuclear reaction rates and conductive opacities are the same as those which we have used.

S01 give numerical results only for the helium core flash phase.
Their value of $\log (L/L_{\sun}) = 2.314$ at the start of the
core helium flash is slightly smaller than our model 10cf value and
their value of $M_{1} \ups{max} = 0.482 \msun$ is also smaller than ours. 
However, they find $M \lows{BCS} = 0.151 \msun$, while our model
10cf gives $M \lows{BCS} = 0.3390 \msun$ and none of our other models
ignite a helium core flash with $M_{\rm BCS}$ smaller than $0.3 \msun$,
except in the case of central ignition (Table~\ref{tab:he-flash}). 
Because of the much larger mass between the base of
the convective shell and the location of the base of the hydrogen-rich layer,
the time required for the outer edge of the convective shell to
reach hydrogen-rich material is much larger in the S01 model than in our models:
$\Delta t \lows{mix} = 10$ yr for S01 versus ${\Delta t}_{\rm mix} 
= 10^{-3}$ - $10^{-2}$ yr for all our models, irrespective of the input
physics. 
  We note that it takes more than 1000 yr for convection generated in the center to reach the maximum extension in mass and ${\Delta t}_{\rm mix}$ for S01 falls in the middle of our two cases.

In an effort to reproduce the S01 results, we constructed a $1.0 \msun$ model 10cf$^\prime$, using the fitting formula for neutrino energy-loss rates given by \citet{mun85} which does not include neutrino bremsstrahlung. 
   The contribution of neutrino loss is rather small as compared with the gravitational energy release in the core, and yet, affects the internal structure of helium core significantly. 
    In Fig.~\ref{fig:str}, we compare the structure line at core mass $M_1 = 0.49 \msun$ in which the neutrino energy loss rate ($L_\nu = 0.70 L_\odot$) is smaller than the rate of gravitational energy release ($L_g = 2.3 L_\odot$) and the helium burning rate is still small ($L_{\rm He} = 0.35 L_\odot$). 
   We see that the maximum temperature shifts to the inner shell as much as $\Delta \log \rho \simeq 0.13$. 
   The reason for this inward shift is that, near the stellar center, the energy-loss rate due to neutrino bremsstrahlung is comparable to the energy-loss rate due to other neutrino processes; 
   neglect of the neutrino bremstrahlung contribution means that the cooling rate in central regions is reduced from what it would otherwise have been. 
   Accordingly, the initiation of a helium-burning thermonuclear runaway occurs for a smaller core mass than would otherwise be the case. 
   We find that, when $M_{1} = 0.4968 \msun$, helium is ignited at a mass point $M \lows{BCS} = 0.2933 \msun$, which is about 10\% smaller than we find for our 10cf model when neutrino bremstrahlung is included. 
   The 10\% reduction we have found is, however, far too small to account for the S01 result but we suspect that differences in neutrino energy-loss rates are in part responsible for the small value of $M \lows{BCS}$ found by S01. 
   Since radiative opacities, conductive opacities, and nuclear reaction rates are presumably also not responsible, differences in the EOS may be another source of the discrepancy. 
   In dense stellar matter, Coulomb corrections reduce the pressure, leading to an increase in the density and to a reduction of core radius.  
   The increase in the gravity entails the higher temperature in the hydrogen burning shell and heats up the core.  
   To explore this point more quantitatively, we examine conditions in the 10cf$^\prime$ model for the same core mass as the S01 model ignites helium ($M_1 = 0.482 \msun$). 
   In Fig.~\ref{fig:str}, we locate by filled circles the density and temperature of the hydrogen-burning shell and the density and temperature at the mass point $M_{r} \approx 0.151 \msun$ when $M_1 = 0.482 \msun$. 
   We conclude that the S01 structure curve must be fairly different from ours in the sense that their internal core might be kept much hotter despite the neutrino energy loss, presumably due to larger Coulomb corrections. 

Finally, we compare the $1 \msun$ model of SLL02 with our model 10nac. 
   SLL02 provide many evolutionary tracks of zero metallicity models covering a large mass range. 
   They do not encounter the He-FDDM phenomenon and their evolutionary calculations for low mass stars extend to the AGB phase.  
   Much of the input physics they adopt is the same as we have used to construct model ``nac''. 
   The conductive opacities, radiative opacities, nuclear reaction rates, and the mixing length parameter ($\alpha = 1.5$) are presumably the same. 
   However, they use a different EOS and a different treatment of the nuclear screening factor \citep{gra73}. 
   They do not comment on the choice of neutrino energy-loss rates. 

With respect to the core He-H flash, the minimum carbon abundance by mass for the appearance of convection at the center in their models is $\log X_{12C} \simeq -11.5$ for various model masses and this agrees well with our results for lower mass models. 
   The main sequence lifetime, measured by the age at the turn-off point, is $t \lows{TO} = 6.56$ Gyr in excellent agreement with our result of $6.53$ Gyr (see Table~\ref{tab:model}); 
   this agreement is to be expected since both opacities and nuclear reaction rates are the same in both cases. 
   For the same reason, the CN-cycle takes over as the main energy-production mechanism at essentially the same abundance of carbon at the center: 
   $X_{c} = \pow{5.8}{-4}$ in the SLL02 model and $\pow{5.71}{-4}$ in ours.
   The maximum mass of the convective core and the maximum helium-burning luminosity are very similar in the two cases: 
   $M \lows{ECS} \simeq 0.095 \msun$ and $L \lows {He} \ups{max} \simeq 10^{-7} L_{\sun}$ in the SLL02 model, compared with $M \lows{ECS} = 0.104 \msun$ and $L \lows {He} \ups{max} \simeq = \pow{2.51}{-7} L_{\sun}$ in ours. 
   At the maximum helium-burning luminosity, element abundances are $X_{12} = \pow{6.50}{-12}$, $X_{14} = \pow{2.19}{-10}$, and $X_{16} = \pow{2.77}{-12}$ in the SLL02 model, compared with $X_{12} = \pow{6.64}{-12}$, $X_{14} = \pow{1.19}{-10}$, and $X_{16} = \pow{8.97}{-13}$ in our model. 

SLL02 do not find a shell He-H flash, consistent with our results. 
   Very similar results are obtained for stellar luminosity and the mass of the helium core at the RGB tip; 
   they find $\log L \ups{tip} = 2.357$ and $M_{1} \ups{tip} = 0.497 \msun$, compared with $\log L \ups{tip} = 2.372$ and $M_{1} \ups{tip} = 0.4922 \msun$ for our model. 
   Core helium burning begins when $M \lows{BCS} = 0.31 \msun$ for SLL02 and $M \lows{BCS} = 0.3320 \msun$ in model 10nac (see Table~\ref{tab:he-flash}). 
   SLL02 actually find that hydrogen mixes into the convective zone driven by core helium flash, but, because the hydrogen abundance that appears in the zone is so small ($X < 10^{-8}$), the luminosity due to hydrogen burning is relatively small ($L \lows{H} \lesssim 10^{4} L_{\sun}$), so they neglect the effects. 
   As mentioned earlier, their result may be an artifact occasioned by their assumption that convective mixing is instantaneous.

For completeness, we discuss additional properties of our $1 M_\odot$ model 10nac and, when possible, compare with properties of the other models. 
At the turn-off point, our model has $M_{1} = 0.2032 \msun$ and
the abundance of hydrogen at the center is finite with the value
$X_{c} = \pow{1.57}{-2}$. 
After 553 Myr of evolution beyond the turnoff point, a convective core
is formed, driven by the He-H core flash. 
Still 3.06 Myr later, the hydrogen-burning luminosity reaches a maximum
of $L \lows{H} \ups{max}=24.6 L_{\sun}$, with $L_{\rm pp}=18.3 L_{\sun}$
and $L_{\rm CN} = 6.29 L_{\sun}$ being, respectively, the contributions of
the pp-chain reactions and of the CN-cycle reactions. 
The blue loop in the H-R diagram is characterized by
$3.795 \leq \log \teff$ (K) $\leq$ 3.863, and 1.27 $\leq \log L_{s} \leq 1.33$. 
During the flash, the convective core grows to a maximum mass of
$M \lows{ECS} = 0.104 \msun$, which is between the masses found by
W00 and SLL02.
The maximum carbon abundance is achieved at nearly the same time that
the convective shell achieves its maximum mass and has the value
$X_{12} \ups{max} = \pow{7.146}{-12}$. 
As evident from Fig.~\ref{fig:rhot}, the inner CN-burning shell passes through
the site of this central convective zone when, at the center, 
$\log \rho_{c} = 5.017$ and $\log T_{c} = 7.801$.
A temperature plateau develops in the outer core, as is characteristic
of all zero metallicity models, and as the core mass continues to increase,
a temperature inversion is formed \citep{fuj84}.
Thereafter, energy transport into the center plays a crucial role
in determining the maximum temperature in the core. 
Our model center evolves into a region of strong degeneracy, reaching the maximum central density $\log \rho_{c} \ups{max} =6.074$ at $\log T_{c} = 7.840$. 
When the helium flash ignites off-center, the core mass is
$M_{1} \ups{max} = 0.5028 \msun$ (see Table~\ref{tab:model}) and
the luminosity is $L \ups{max} = 251.6 L_\odot$. 
At maximum helium luminosity, the central abundances of CNO elements are
$X_{12}=\pow{1.18}{-5}$, $X_{14}=\pow{5.35}{-10}$, and $X_{16}=\pow{9.21}{-8}$.
In our model, $T_{c}=\pow{5.622}{7}$ K and  $\rho_{c}=\pow{8.498}{5}$ g cm$^{-3}$ at this time.

\subsection{Influence of \tal\ rates}}\label{sec:tal}

In this subsection, we discuss the relevance of characteristics of \tal\ rates to the evolution of zero-metallicity stars.  
  Recently, the properties of resonances other than the so-called Hoyle resonance in \nuc{12}C and their effects on \tal\ rates have been discussed experimentally and theoretically since the compilation of NACRE \citep[see for example,][]{ito04b,fyn05,kur05}.  
   On the other hand, the non-resonant term of \tal\ rate may also be subject to uncertainty in the cross section because the nuclear theory hardly determines the behavior of three body interactions (K. Kat$\bar{\rm o}$ 2007, private communication). 
   Considering these improvement in the field of nuclear physics, it is important to examine the effects of resonant and non-resonant terms of \tal s, separately.  
   Indeed, for $Z=0$ models, the non-resonant term of \tal\ rate can change the evolutionary behavior drastically at the hydrogen shell burning phase.

As a test for exploring the contribution of non-resonant term, we compute the models of 1.0 and $1.1 \msun$ by adopting the nuclear reaction rates of \citet{fow75} (hereafter FCZ75) with the same other input physics as the models in this work. 
   The main difference of FCZ75 rates from the NACRE rates is the consideration of non-resonant term in \tal.  
   Since FCZ75 have not yet taken into account the effect, the \tal\ rate drops by far more rapidly than the NACRE rates below $\log T \lesssim 7.89$. 
   Other cross sections may differ by within factor of 2 or 3 \citep{sud03} and do not affect the qualitative results. 

\begin{figure}
\plotone{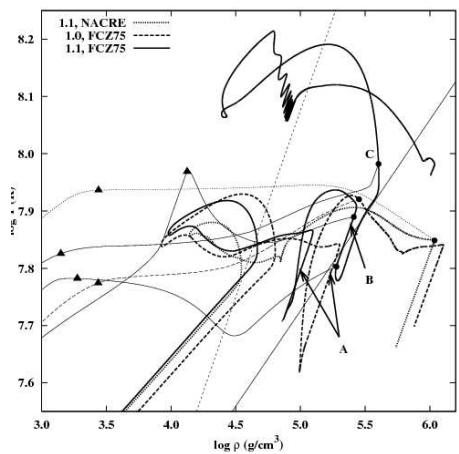}
\caption{
Evolutionary tracks (thick lines) and structure at some stages (thin lines)
in the temperature-density plane for selected models.
The labels on top left corner denote the mass and the adopted nuclear
reaction rates for each model.
Circles and triangles denote the center
of each model and the hydrogen-burning shell where the hydrogen-burning rate
is maximum.
The solid straight line and the dashed straight line denote constant
$\Gamma = 0.5$ and $T = 0.3 T_{F}$, respectively, where $\Gamma$ is
the Coulomb coupling constant and $T_{F}$ is the Fermi temperature.
See text for detail on labels ``A'' to ``C''.
}
\label{fig:shellHeH}
\end{figure}

Figures~\ref{fig:shellHeH} shows the evolutionary tracks and interior structures in the density-temperature plane for these models with the FCZ75 rates (refereed as models 10fcz and 11fcz, respectively, in the following) and compare them with those of our models with the NACRE rates.  
   The effect of different nuclear reaction rates is apparent before the depletion of hydrogen in the center. 
   Both models with and without the non-resonant term experience the core He-H flash, the FCZ models postpone it until higher central temperature than the NACRE models; 
   in the latter models, it is ignited at the low temperatures where the non-resonant term is effective ($\log T < 7.89$). 
   Because of stronger electron degeneracy, the FCZ75 models undergo much stronger flashes ($L_{\rm He} = 1.06 \times 10^4$ and $78 L_\odot$ for models 10fcz and 11fcz, respectively), as seen from larger loops in this figure, than our models 10nac and 11nuc ($L_{\rm He} = 24$ and $29 L_\odot$, respectively), which entails larger extension of flash convection; 
   for models 10fcz and 11fcz, the flash convection reaches to the shells of $M_{\rm conv} = 0.181$ and $0.149 \msun$ at the maximum extension, respectively, about twice as large as compared to our models 10nac and 11nac, with the CN abundance of $X_{\rm CN} = 6.0 \times 10^{-10}$ and $ 2.6 \times 10^{-10}$, respectively. 

There is also a remarkable difference in the evolutionary tracks during the hydrogen-shell burning between the models with and without the non-resonant terms. 
   In the hydrogen burning-shell, the CN-burning with carbon produced by \tal\ contributes considerably to the total energy generation though the contribution to the total energy is still smaller than that of p-p chain reactions for the stars of mass $1.2 > M/ \msun > 0.9 $, because of lower entropy in the hydrogen burning shell as compared with the stars of younger populations. 
   In the models without the non-resonant term, in particular, there form two local maxima in the energy generation rates, a narrow one associated with CN burning and localized at the base of small hydrogen abundance ($X < 0.01$), and a broad one associated with pp-chain reactions in the middle of large hydrogen abundance (nearly half of the surface hydrogen abundance). 
   When the hydrogen burning shell passes across the inner sphere, occupied by the convection during the core He-H flash, a thermonuclear runaway is triggered at the base of hydrogen-burning shell under electron degeneracy, which is a similar situation as the ignition of a core He-H flash in the center; 
   in addition, at the base of hydrogen burning shell, the pressure scale height is much smaller than the radial distance to the center, i.e., $V = r / \vert d r / d \ln P \vert = G M_r \rho / r P \gg 1$, and the flat configuration also contributes to the instability \citep{sug78}. 
   It occurs in the shell of mass $M_r = 0.212 \msun$ with $M_1 = 0.340 \msun$ for mode 10fcz and $M_r = 0.170 \msun$ with $M_1 = 0.352 \msun$ for the first shell flash of model 11fcz.

On the other hand, the models with the non-resonant term stably burn hydrogen, although electrons are degenerate ($\Psi \simeq 8$) at the base of the hydrogen burning shell. 
   This different behavior stems mainly from the difference in the temperature dependence of the resonant and non-resonant terms, rather than from the difference in the burning rate itself, as can be seen from the analysis of thin shell burning by \citet[see also Fujimoto 1982]{sch65}. 
   At temperatures of $\log T < 7.9$, the \tal\ rates by FCZ75 and NACRE give a large difference in their temperature dependences when compared at the same burning rates. 
   For less massive FCZ75 models of $M \le 0.9 \msun$, the entropy is smaller and hydrogen is burnt before carbon production becomes appreciable. 
   For FCZ75 models of mass $M \ge 1.2 \msun$, higher temperature as well as weaker electron degeneracy tend to stabilize hydrogen shell burning; 
   in addition, as the contribution of CN burning with products of \tal\ reaction as catalysis overweighs the p-p chain reactions, the pressure scaleheight grows comparable to the radial distance to the center, whicl also stabilize the shell burning by making the heat capacity negative with hydrostatic readjustment \citep[e.g., see ][]{fuj82}.  
   It is noted that the temperature dependence of resonant reaction rate decreases with increase in the temperature, which also stabilizes the hydrogen shell-burning in combination with the reduction in electron degeneracy when the core grows more massive than $M_1 \gtrsim 0.4 \msun$, as is the case for 10fcz model.  

During the shell flash, the CN-cycle burning shell expands, and the hydrogen exhausted core also expands due to the reduction in the weight of overlying layers; 
   the central temperature and density decrease almost adiabatically (Label ``A'' in Fig.~\ref{fig:shellHeH}). 
  During the decay phase of the first shell flash, after the flash-driven convective zone disappears, the core is heated by the flow of energy from the burning shell. 
   For model 11fcz, the second shell flash is ignited at the shell of mass $M_r = 0.287 \msun$ with $M_1 = 0.374 \msun$, when the hydrogen shell-burning passes across the shells, incorporated into the convection and enriched in CN elements during the first shell flash. 
   This flash grows so strong as to drive the convection deep into the hydrogen-rich envelope up to the shell of $M_1 = 0.404 \msun$ (Label ``B''), which expedites the growth of helium core to cause the increase in central density.  
  Since the matter in the convective zone driven by the shell flash is enriched in CNO elements, the CN-cycle reactions dominate the energy generation rate while the burning shell traverse in the site of convective zone. 
   During this stable burning phase, the growth rate of core is large because of small hydrogen abundance therein. 
   Accordingly, concomitant rapid compression of helium core increases the central temperature. 
   This enhanced growth of helium core after the shell flashes leads directly to the ignition at the center of a helium core flash (Label ``C'') with a small core masses.  
  This is the case if the convective zone is sufficiently large, as in the case for the second shell flashes of 11fcz model, while it is barely missed in 10fcz model.  

The most important point by this test is that the contribution of the non-resonant term is discernible in the circumstance of $Z=0$, although it may be difficult to detect differences by the observations. 
   There is a difference in the mass range that the off-center helium core flashes occur. 
   If the non-resonant term is included in the \tal\ rate, the central helium burning occurs at $\geq 1.2 \msun$. 
   Otherwise, it occurs at $\geq 1.0 \msun$ if we use the smaller conductivity by Iben's approximates to \citet{hub69} and by \citet{can70} and the neutrino loss rates by \citet{bea67} as in FII00, both slightly smaller than ours. 
   Since the lifetime of $1.0 \msun$ model is $\sim$ 7 Gyr (Table~\ref{tab:model}), such stars cannot be seen in the present halo if they were born. 
   Only clue to those objects will be binary mass transfer between giants of mass $\ge1.0 \msun$ and dwarfs of mass $\le 0.8 \msun$. 
   Since FII00 predict the abundance ratio of C/N $\sim 1$ for He-FDDM at RGB, while C/N $\gtrsim 5$ for He-FDDM at AGB, it will be crucial for the constraint on the estimate of nuclear reaction rate to determine the abundance ratio of C/N and the mass of the primary for $Z=0$.

\section{Conclusions}

We have explored the evolution of low-mass, zero-metallicity stars with the most recent input physics.   
  The mass range is $0.8 \msun \leq M \leq 1.2 \msun$ in step of $0.1 \msun$ and the initial composition is $X = 0.767$, $Y=0.233$, and $Z=0$.  
  Calculations extend from the zero-age main sequence to the beginning of hydrogen mixing into the helium convective region on the RGB or at the start of the TPAGB phase, depending on the characteristics of the helium core flash.

\begin{enumerate}

\item The emergence of CN-cycle reactions as important contributors to nuclear energy production occurs during the core hydrogen-burning phase in models of mass $M \geq 0.9 \msun$ in consequence of the formation of carbon by the 3$\alpha$ reaction. This phenomenon is independent of the adopted physics and its importance is a function only of the initial mass and metallicity. 

\item The models of $M \leq 1.1 \msun$ undergo an off-center helium flash and hydrogen-mixing into helium flash-convection, leading to helium-flash driven deep mixing at the tip of red giant branch.  
   On the other hand, the models of mass $M \ge 1.2\msun$ ignite the core helium flash at the center and postpone the He-FDDM until the helium shell flashes occur during the early phase of thermal pulsation at the asymptotic giant branch. 
   For models of mass $0.8 \msun$ and $0.9 \msun$, our results coincide qualitatively with those first found by \cite{fuj90} and \citet{fuj00}. 
   For $1.0 \msun \leq M \leq 1.1 \msun$, whether or not a helium core flash is ignited off the center and hydrogen is mixed inward into the convective zone driven by it depend on the adopted nuclear reaction rates.   
\end{enumerate}

We have also compared our results with those of other investigations. Our models made with the most up-to-date input physics agree well with the models of \citet{pic04}, \citet{wei00} and \cite{sie02} during evolution on the main sequence and the RGB. 
   In particular, we obtain nearly the same results as do \cite{sie02}, although we do not follow evolution after the mixing of hydrogen into the helium flash driven convective zone at the beginning of the helium core flash. 
   To check the behavior of the He-FDDM event, we need to treat mixing with a time-dependent algorithm. 
   On the other hand, even after adopting the same radiative opacities, conductive opacities, nuclear reaction rates, and neutrino energy-loss rates as \cite{sch01}, we are not able to obtain the inner ignition at the onset of core helium burning which they find; 
  we suspect that the discrepancy may be due to differences in the EOS for the Coulomb corrections in the liquid and solid states and due to the neglect in their work of neutrino energy losses associated with neutrino bremsstrahlung.

By treating the resonant and non-resonant rates of \tal s separately, we demonstrate that the non-resonant term plays a critical role in the low-mass, zero-metal stars. 
   The neglect of non-resonant term causes the lower border in mass of central helium burning, i.e., $\geq 1.0 \msun$. 
   This explains the discrepancy of the results between ours and \citet{fuj00}, which stems mainly from the difference in the temperature dependence rather than in the energy generation rates themselves.  
   We first point out the possibility of discerning the effect of non-resonant term of \tal\ from the evolution of stars other than at low temperature regime in the accreting degenerate stars \citep{nom85}.  
   It is important to precisely determine the abundances and the properties of extremely metal-poor stars to constrain the nuclear reaction rates. 

\acknowledgments

We are grateful to I. Iben Jr. for improving and revising our manuscript. 
   We wish to thank A. Ohnishi and K. Kat$\bar{\rm o}$ for valuable comments on uncertainties on nuclear reaction rates. 
   This work is part of a PhD. thesis constructed at Hokkaido University and is in part supported by a Grant-in-Aid for Science Research from the Japanese Society for the Promotion of Science (15204010, 18104003).   

\appendix

\section{Conductive Opacities}

The most up-to-date conductive opacities are those of
\citet[hereinafter I83]{ito83}.
We consider an ion mixture
consisting of $n$ species of nuclei. The conductive opacity
$\kappa_{c}$ for temperature $T_{8}$ (units of $10^{8}$ K)
and density $\rho_{6}$ (units of $10^{6}$ g cm$^{-3}$) is
taken from eq. (7) in I83,
\begin{equation}
\kappa_{c} = \pow{1.280}{-3} \left( \sumin X_{i} A_{i}
 \left\langle S \right\rangle_{i} \right) \left[ 1 +
 1.018 \left( \sumin \frac{Z_{i}}{A_{i}} X_{i} \right)^{2/3}
 \rho_{6}^{2/3} \right] \left( \frac{T_{8}}{\rho_{6}}
 \right)^{2} \left[ \textrm{cm}^{2} \textrm{g}^{-1} \right],
\end{equation}
where $\langle \rangle$ denotes the average over the nuclear
species and $Z_{i}$, $X_{i}$, and $A_{i}$ are the atomic number,
the mass fraction, and the atomic mass number of $i^{\rm th}$
nucleus. The average value is taken as
\begin{eqnarray}
\left\langle S \right\rangle_{i} &=& \ave{S_{-1}}_{i} -
 \frac{1.018 \left( \sum Z_{i} X_{i} / A_{i} \right)^{2/3}
 \rho_{6}^{2/3}}{1 + 1.018 \left( \sum Z_{i} X_{i} / A_{i}
 \right)^{2/3} \rho_{6}^{2/3}} \ave{S_{+1}}_{i}. 
\end{eqnarray}
The quantities $\ave{S_{-1}}_{i}$ and $\ave{S_{+1}}_{i}$ are
calculated with eqs. (8) and (9) in I83 and with the parameters
according to eq.(19) of \citet{ito04}:
\begin{eqnarray}
\gumi &=& \displaystyle \frac{Z_{i}^{5/3} e^{2}}{a_{e} k_{B} T}
 = 0.2275 \frac{Z_{i}^{5/3}}{T_{8}}
 \left( \rho_{6} \sumjn \frac{X_{j} Z_{j}}{A_{j}} \right)^{1/3} \\
 x_{i} &=& 0.45641 \ln \Gamma_{i} - 1.31636 \\
 r_{s} &=& \pow{1.388}{-2} \left( \sumjn 
\frac{Z_{j}}{A_{j}} X_{j} \rho_{6} \right)^{-1/3}, 
\end{eqnarray}
where \gumi \ is the Coulomb coupling constant for the $i^{\rm th}$
nucleus, $a_{e}$ the electron-sphere radius defined as
$a_{e} = (3 / 4 \pi n_{e})^{1/3}$ with the electron number density $n_{e}$,
and $r_{s}$ the electron density parameter.

The expressions for various mixtures are to be be considered
as first approximations, compared with the formulae in I83
which are for pure compositions and are accurate solutions.
In particular, the approximation for a mixture of elements
with very different $Z$'s (e.g., $\nucm{1}{H}$ and $\nucm{56}{Fe}$)
is not very accurate; however, the exact solution for such
mixtures is not presently attainable. For mixtures of elements
of comparable $Z$'s (e.g., $\nucm{12}{C}$ and $\nucm{16}{O}$),
the approximation is fairly accurate.

The I83 results are strictly applicable only for
$T \leq T_{F}$, where $T_{F}$ is the Fermi
temperature defined by eq. (1) in I83.
In this work, $\kappa_{c}$ as defined by I83
and $\kappa_{c}$ as defined by \citet{ibe75} are interpolated
with one other over the range $0.5 \leq T / T_{F} \leq 2.0$
using a sin squared algorithm.

Another algorithm is used to extrapolate I83 results for values of
$\Gamma$ outside the region defined by $2 \leq \Gamma \leq 160$.
Both $\ave{S_{-1}}$ and $\ave{S_{+1}}$ are constrained by
$\ave{S_{-1}} \leq \ave{S_{-1}}_{\rm lim}$ and $\ave{S_{+1}} \leq
\ave{S_{+1}}_{\rm lim}$, where the upper limits are calculated by
demanding that, in eq. (6) in I83,
\begin{eqnarray}
S \left( \frac{k}{2 k_{F}} \right) &=& 1, \\
\epsilon \left( \frac{k}{2 k_{F}} , 0 \right) &=& \displaystyle
 \frac{k^{2} + k_{TF}^{2}}{k^{2}}. 
\end{eqnarray}
where $k_{\rm TF}$ is the Thomas-Fermi wavenumber. The first condition
assumes that interactions between ions can be neglected and the second
condition expresses the Thomas-Fermi approximation for electron screening.
These conditions give for the upper limits:
\begin{eqnarray}
\ave{S_{-1}}_{lim} &=& \frac{1}{2} \left[ \ln \left( 1 +
 \frac{4 k_{F}^{2}}{k_{TF}^{2}} \right) -\left(1+\thomfer\right)^{-1}
 \right], \label{eq:sm1} \\
\ave{S_{+1}}_{lim} &=& \frac{1}{2} - \thomfer
 \left[ \ln \left(1+ \frac{4 k_{F}^{2}}{k_{TF}^{2}} \right)
 + \frac{1}{2} - \left(1 + \thomfer\right)^{-1} \right] +
 \frac{1}{2} \left( \thomfer \right)^{2}
 \left(1 + \thomfer\right)^{-1} \label{eq:sp1} 
\end{eqnarray}
The factor $k_{TF}^{2} / (4 k_{F}^{2})$ is given by
\begin{eqnarray}
\thomfer &=& \frac{\alpha}{\pi} \left( 1 + \frac{1}{b^{2}} \right)^{1/2}, \\
\alpha &=& \frac{1}{137}, \\
b &=& \frac{\hbar k_{F}}{m c} = \frac{1}{137} \left( \frac{9 \pi}{4} \right)^{1/3} r_{s}^{-1} 
\end{eqnarray}
where $\alpha$ is the fine structure constant and $b$ is taken from I83.
This holds both for relativistic degeneracy ($\rho_{6} \geq 1$) and
for non-relativistic degeneracy ($\rho_{6} \leq 1$).

The interpolation between I83 and \citet{hub69} as approximated by
\citet{ibe75} is shown in Fig.~\ref{fig:cop}.
In the actual computations, the upper limits \ref{eq:sm1} and \ref{eq:sp1}
are infrequently invoked.

\section{Neutrino Losses By Bremsstrahlung}

We describe here the estimate of the neutrino energy-loss rate due to
bremsstrahlung for the case of an ion mixture as discussed by
\citet[hereinafter I96]{ito96}. For any nuclear species, gas,
liquid, and solid states for ions are taken into consideration.
The total energy-loss rate is given by
\begin{equation}
\breq{brems} = {\sum_{i}} X_{i} \left( \breq{gas}^{i} +
 \breq{liq}^{i} + \breq{sol}^{i} \right) ,
\end{equation}
where the $X_{i}$ are nuclear abundances and the summation is
taken over the species $\nucm{4}{He}$, $\nucm{12}{C}$, and $\nucm{16}{O}$.
The gas state corresponds to the scheme for weakly degenerate electrons
 and $\breq{gas}^{i}$ is based on eq. $(5.1)$ in I96:
\begin{equation}
\breq{gas}^{i} = 0.5738 \, \frac{Z_{i}^{2}}{A_{i}} T_{8}^{6}
 \rho \left( C_{+} \bref{gas} - C_{-} \breg{gas} \right), 
\end{equation}
where
\begin{eqnarray}
C_{+} = \frac{1}{2} \left\{ \left( C^{2}_{V} + C^{2}_{A} \right) + n \left( C^{\prime 2}_{V} + C^{\prime 2}_{A} \right) \right\}, \\
C_{-} = \frac{1}{2} \left\{ \left( C^{2}_{V} - C^{2}_{A} \right) + n \left( C^{\prime 2}_{V} - C^{\prime 2}_{A} \right) \right\}. 
\end{eqnarray}
All values of the right hand side of these equations are defined in I96.
$\bref{gas}$ and $\breg{gas}$ are independent on the nuclear species other than the number of electrons. 
$\bref{gas}$ is given by eq. $(5.2)$ in I96, but the parameter $\eta$
is given by
\begin{equation}
\eta = \muii \left( \pow{7.05}{6} \, T_{8}^{1.5} + \pow{5.12}{4}
  \, T_{8}^{3} \right)^{-1}.
\end{equation}
$\breg{gas}$ is given as follows according to eq. $(5.9)$ in I96:
\begin{eqnarray}
\breg{gas} = \left[\left( 1 + 10^{-9} \mui \right)
 \left( a_{3} + a_{4} T_{8}^{-2} + a_{5} T_{8}^{-5}
 \right) \right]^{-1} \nonumber \\
+ \left[b_{3} \left( \mui \right)^{-1} +
 b_{4} + b_{5} \left( \mui \right)^{0.656} \right]^{-1}
\end{eqnarray}
According to I96, the applicable range of $\breq{gas}^{i}$ is
determined by $T > 0.01 \, T_{F}$ for $\nucm{4}{He}$ and $T > 0.3 \,
 T_{F}$ for $\nucm{12}{C}$ and $\nucm{16}{O}$,
 where $T_{F}$ is the Fermi temperature:
\begin{equation}
T_{F} = \pow{5.9302}{9} \, \left\{ \sqrt{1 + 1.018 
\left( \sum \frac{X_{i} Z_{i}}{A_{i}} \rho_{6} \right)^{2/3} }
 - 1 \right\} \quad \left[ \textrm{K} \right]. 
\end{equation}

$\breq{liq}^{i}$ is calculated according to \S 5.2 in I96.
For $T < 0.3 \, T_{F}$ (or $T < 0.01 \, T_{F}$ for $\nucm{4}{He}$),
the formula for the bremsstrahlung process in the liquid state
is adopted if the parameter, which considers ion-ion correlation
effect according to \citet{ito04},
\begin{equation}
\gumi = 0.2275 \, \frac{Z_{i}^{5/3}}{T_{8}}
 \left( \sumjn \frac{X_{j} Z_{j}}{A_{j}} \rho_{6} \right)^{1/3}
\end{equation}
satisfies $\gumi < 180$.
Here, $\breq{liq}^{i}$ is given in cgs units as
\begin{equation}
\breq{liq}^{i} = 0.5738 \, \frac{Z_{i}^{2}}{A_{i}}
 T_{8}^{6} \rho \left( C_{+} \bref{liq} - C_{-} \breg{liq} \right), 
\end{equation}
where \bref{liq} and \breg{liq} are provided by eqs. $(5.19)$ and
$(5.20)$, respectively, in I96. These quantities are functions
of $u$ and \gumi \ via
\begin{eqnarray}
v_{i} &=& \summ \alpha_{m} \gumi^{- \frac{m}{3}}, \\
w_{i} &=& \summ \beta_{m} \gumi^{- \frac{m}{3}}.
\end{eqnarray}

For $\gumi \geq 180$, we assume for the bremsstrahlung process that
\begin{equation}
\breq{sol}^{i} = \breq{lattice}^{i} + \breq{phonon}^{i} 
\end{equation}
where, according to eq. $(5.29)$ in I96,
\begin{eqnarray}
\breq{lattice}^{i} &=& 0.5738 \, \frac{Z_{i}^{2}}{A_{i}}
 T_{8}^{6} \rho \left( C_{+} \bref{lattice} - C_{-} \breg{lattice} \right)
 \label{eq:b13}, \\
\breq{phonon}^{i} &=& 0.5738 \, \frac{Z_{i}^{2}}{A_{i}}
 T_{8}^{6} \rho \left( C_{+} \bref{phonon} - C_{-} \breg{phonon}
 \right) \label{eq:b14}.
\end{eqnarray}
Note that we adopt $f_{\textrm{band}} = 1$ in $\breq{lattice}^{i}$.
Each factor such as $F$ and $G$ in the right-hand side of \ref{eq:b13}
 and \ref{eq:b14} depends on the nuclei via the parameters:
\begin{eqnarray}
v_{i} &= \summ \alpha_{m} \gamim \hspace{5mm} w_{i} &= \summ \beta_{m}
 \gamim \\
v_{i}^{\prime} &= \summ \alpha_{m}^{\prime} \gamim \hspace{5mm}
 w_{i}^{\prime} &= \summ \beta_{m}^{\prime} \gamim .
\end{eqnarray}

\begin{table}
  \begin{center}
    \begin{tabular}{lcccccrr}
      \hline
      \hline
      Model  &  $M$  &  $\log \teff \ups{TO}$  &  $\log L_{s} \ups{TO}$   &
      $\log \teff \ups{tip}$  &  $\log L_{s} \ups{tip}$  &  $t \lows{TO}$  &  $t \lows{tip}$   \\
             & ($\msun$)         & (K)                     &  ($M_{\odot}$)           &
      (K)                     &  ($L_{\odot}$)           &  (Gyr)          &  (Gyr)               \\
      \hline
      08nac & 0.8 & 3.833  & 0.4221 & 3.697  & 2.425 & 14.03 & 15.84 \\ 
      09nac & 0.9 & 3.866  & 0.6559 & 3.698  & 2.411 & 9.31  & 10.45 \\ 
      10nac & 1.0 & 3.909  & 0.8589 & 3.700  & 2.401 & 6.53  &  7.28 \\ 
      11nac & 1.1 & 3.952  & 1.0403 & 3.703  & 2.369 & 4.77  &  5.28 \\ 
      12nac & 1.2 & 3.987  & 1.1991 & 3.858  & 1.870 & 3.76  &  4.17 \\ 
      \hline
      08cf  & 0.8 & 3.831  & 0.4126 & 3.696  & 2.446 & 14.03 & 15.88  \\ 
      09cf  & 0.9 & 3.864  & 0.6589 & 3.697  & 2.438 & 9.35  & 10.47 \\ 
      10cf  & 1.0 & 3.908  & 0.8613 & 3.700  & 2.413 & 6.54  &  7.28  \\
      11cf  & 1.1 & 3.949  & 1.0398 & 3.703  & 2.386 & 4.77  &  5.28 \\ 
      12cf  & 1.2 & 3.986  & 1.2086 & 3.852  & 1.872 & 3.61  &  4.00 \\ 
      \hline
    \end{tabular}
    \caption[]{Evolutionary Characteristics of Low-Mass, Population III Models with Up-to-Date Input Physics}
    \label{tab:model}
  \end{center}
\end{table}

\begin{table}
  \begin{center}
    \begin{tabular}{lccccccc}
      \hline
      \hline
      Model  & $M_{1} \ups{max}$ & $M \lows{BCS}$ & $\lhe \ups{max}$  & $T \lows{BCS} \ups{max}$ & $\lhe \ups{mix}$ &  $\Delta t^{\prime}$ & $\Delta t \lows{mix}$ \\
            & ($\msun$) & ($\msun$) & ($L_{\odot}$)  & (10$^{6}$K) & ($L_{\odot}$) &(yr)& (yr) \\
      \hline
      08nac  & 0.5113 & 0.3579 & 9.913  & 244.5 & 9.263 & 750 & \pow{6.07}{-3} \\
      09nac  & 0.5078 & 0.3480 & 9.876  & 241.1 & 9.166 & 813 & \pow{7.51}{-3} \\
      10nac  & 0.5028 & 0.3320 & 9.834 & 237.1 & 9.038 & 846 & \pow{9.81}{-3} \\
      11nac  & 0.4929 & 0.2986 & 9.722 & 226.9 & 8.744 & 953 & \pow{1.82}{-2} \\
      12nac  & 0.3840 & 0.0     & 3.829 & 126.8  & -  & -     & -              \\
      \hline
      08cf   & 0.5151 & 0.3717 & 9.882 & 248.6 & 9.264 & 715 & \pow{5.91}{-3} \\
      09cf   & 0.5121 & 0.3629 & 9.858 & 245.3 & 9.231 & 749 & \pow{6.49}{-3} \\
      10cf   & 0.5054 & 0.3390 & 9.831 & 239.4 & 9.026 & 858 & \pow{9.99}{-3} \\
      11cf  & 0.4966  & 0.3115 & 9.728 & 231.1 & 8.776 & 941 & \pow{1.69}{-2} \\
      12cf  & 0.3829  & 0.0     & 3.393   & 124.2  & -  & -     & -              \\
      \hline
    \end{tabular}
    \caption{Characteristics of Helium Core Flash of Model Stars with Up-to-Date Input Physics}
    \label{tab:he-flash}
  \end{center}
\end{table}

\clearpage

\begin{thebibliography}{}
\bibitem[Abe(1959)]{abe59} Abe, R. 1959, Prog. Theor. Phys., 22, 213
\bibitem[Alexander \& Ferguson(1994)]{ale94} Alexander, D. R. \& Ferguson, J. W. 1994, \apj, 437, 879
\bibitem[Angulo et al.(1999)]{ang99} Angulo et al. 1999, Nucl. Phys. A, 656, 3
\bibitem[Austin, Trentelman \& Kashy(1971)]{aus71} Austin, S.M, Trentelman, G.F. \& Kashy, E. 1971, \apjl, 163,L79
\bibitem[Beaudet et al.(1967)]{bea67} Beaudet, G., Petrosian, V., \& Salpeter, E. E. 1967, \apj, 150, 979
\bibitem[B\"ohm-Vitense(1958)]{bohm58} B\"ohm-Vitense, E., 1958, Zs. f.
Ap., 46, 108
\bibitem[Bowers \& Salpeter(1960)]{bow60} Bowers, D. L. \& Salpeter, E. E. 1960, Physical Review Series, 119, 1180
\bibitem[Caughlan \& Fowler(1988)]{cau88} Caughlan, G. R., \& Fowler, A. W. 1988, At. Data Nucl. Data Tables, 40, 283
\bibitem[Cameron(1959)]{cam59} Cameron, A. G. W. 1959, \apj, 130, 916
\bibitem[Canuto(1970)]{can70} Canuto, V. 1970, \apj, 159, 641
\bibitem[Carr(1961)]{car61} Carr, W. J. 1971, Physical Review Series, 122, 1437
\bibitem[Cassisi \& Castellani(1993)]{cas93} Cassisi, S. \& Castellani, V. 1993, \apjs, 88, 509
\bibitem[Cassisi et al.(1996)]{cas96} Cassisi, S., Castellani, V., \& Tornamb\`e, A. 1996, \apj, 459, 298
\bibitem[Castellani \& Paolicchi(1975)]{cas75} Castellani, V. \& Paolicchi, P. 1975, \apss, 35, 185
\bibitem[Christlieb et al.(2002)]{chr02} Christlieb, N. et al. 2002, \nat, 419, 904
\bibitem[Christy(1966)]{chr66} Christy, R. F. 1966, \apj, 144, 108
\bibitem[Clayton(1968)]{cla68} Clayton, D. D. 1968, in Principles of Stellar Evolution and Nucleosynthesis, (USA: McGRAW-HILL book company), 357
\bibitem[Cohen \& Keffer(1955)]{coh55} Cohen, M. H., \& Keffer, F. 1955, Physical Review Series, 99, 1128
\bibitem[Cox \& Giuli(1968)]{cox68} Cox, A. N., \& Giuli, P. T. 1968, in Principles of Stellar Evolution, Vol.1 (New York: Gordon and Breach), 281
\bibitem[Cox \& Stewart(1970a)]{cox70a} Cox, A. N., \& Stewart, J. N. 1970, \apjs, 19, 243
\bibitem[Cox \& Stewart(1970b)]{cox70b} Cox, A. N., \& Stewart, J. N. 1970, \apjs, 19, 261
\bibitem[D'Antona (1982)]{dan82} D'Antona, F. 1982, \aap, 115, L1
\bibitem[Dicus et al.(1976)]{dic76} Dicus, D. A., Kolb, E. D., Schramm, D. N., \& Tubbs, D. L. 1976, \apj, 210, 481
\bibitem[Esposito et al.(2003)]{esp03} Esposito, S., Mangano, G., Miele, G., Picardi, I., \& Pisanti, O. 2003, Nuclear Physics B, 658, 217
\bibitem[Ezer \& Cameron(1971)]{eze71} Ezer, D. \& Cameron, A. G. W. 1971, \apss, 14, 399
\bibitem[Festa \& Ruderman(1969)]{fes69} Festa, G. G., \& Ruderman, M. A. 1969, Physical Review Series, 180, 1227
\bibitem[Fowler et al.(1975)]{fow75} Fowler, W. A., Caughlan, G. R., \& Zimmerman, B. A. 1975, \araa, 13, 69
\bibitem[Frebel et al.(2005)]{fre05} Frebel, A., Aoki, W., Christlieb, N. et al. 2005, Nature, in press.
\bibitem[Fujimoto(1977)]{fuj77} Fujimoto, M. Y. 1977, \pasj, 29, 331
\bibitem[Fujimoto(1982)]{fuj82} Fujimoto, M. Y. 1982, \apj, 257, 767
\bibitem[Fujimoto et al.(1984)]{fuj84} Fujimoto, M. Y., Hanawa, T., Iben, I. Jr., \& Richardson, M. B. 1984, \apj, 278, 813

\bibitem[Fujimoto et al.(1990)]{fuj90} Fujimoto, M. Y., Iben, I. Jr., \& Hollowell, D. 1990, \apj, 349, 298
\bibitem[Fujimoto et al.(1995)]{fuj95} Fujimoto, M. Y., Sugiyama, K., Hollowell, D., \& Iben, I. Jr. 1995, \apj, 444, 175
\bibitem[Fujimoto et al.(2000)]{fuj00} Fujimoto, M. Y., Ikeda, Y., \& Iben, I. Jr. 2000, \apjl, 529, L25
\bibitem[Fynbo et al.(2005)]{fyn05} Fynbo, H. O. et al. 2005, \nat, 433, 136
\bibitem[Graboske et al.(1973)]{gra73} Graboske, H. C., DeWitt, H. E., Grossman, A. S., \& Cooper, M. S. 1973, \apj, 181, 457
\bibitem[Guenther \& Demarque(1983)]{gue83} Guenther, D. B. \& Demarque, P. 1983, \aap, 118, 262
\bibitem[Hansen(1973)]{han73} Hansen, J. P. 1973, \pra, 8, 3096
\bibitem[Hansen \& Mazighi(1978)]{han78} Hansen, J. P. \& Mazighi, R. 1978, \pra, 18, 1282
\bibitem[Hubbard \& Lampe(1969)]{hub69} Hubbard, W. B., \& Lampe, M. 1969, \apjs, 18, 297
\bibitem[Hollowell et al.(1990)]{hol90}Hollowell, D., Iben, I. Jr., \& Fujimoto, M. Y. 1990, \apj, 351, 245
\bibitem[Iben(1965)]{ibe65}Iben, I. Jr. 1965, \apj, 141, 993
\bibitem[Iben(1975)]{ibe75}Iben, I. Jr. 1975, \apj, 196, 525
\bibitem[Iben \& Tutukov(1984)]{ibe84}Iben, I. Jr., Tutukov, A., V. 1984, \apj, 282, 615
\bibitem[Iben et al.(1992)]{ibe92}Iben, I. Jr., Fujimoto, M., Y., \& MacDonald, J. 1992, \apj, 388, 521
\bibitem[Iglesias et al.(1992)]{igl92}Iglesias, C. A., Rogers, F. J., \& Wilson, B. G. 1992, \apj, 397, 717
\bibitem[Iglesias \& Rogers(1996)]{igl96}Iglesias, C. A. \& Rogers, F. J. 1996, \apj, 464, 943
\bibitem[Itoh et al.(1983)]{ito83}Itoh, N., Mitake, S., Iyetomi, H., \& Ichimaru, S. 1983, \apj, 273, 774
\bibitem[Itoh et al.(1984)]{ito84}Itoh, N., Kohyama, Y., Matsumoto, N., \& Seki, M. 1984, \apj, 280, 787
\bibitem[Itoh et al.(1996)]{ito96}Itoh, N., Hayashi, H., Nishikawa, A., \& Kohyama, Y. 1996, \apjs, 102, 411
\bibitem[Itoh et al.(2004)]{ito04}Itoh, N., Asahara, R., Tomizawa, N., Wanajo, S., \& Nozawa, S. 2004, \apj, 611, 1041
\bibitem[Itoh et al.(2004)]{ito04b}Itoh, M. et al. 2004, Nuclear Physics A, 738, 268
\bibitem[Iyetomi \& Ichimaru(1982)]{iye82}Iyetomi, H. \& Ichimaru, S. 1982, \pra, 25, 2434
\bibitem[Kippenhahn \& Weigert(1990)]{kip90} Kippenhahn, R. \& Weigert, A. 1990, Stellar Structure and Evolution (Berlin : Springer)
\bibitem[Komiya et al.(2007)]{Komiya07}Komiya, Y., Suda, T., Minaguchi, H. Shigeyama, T., Aoki, W., \& Fujimoto, M. Y., 2007, \apj, 658, 367
\bibitem[Kurokawa \& Kat$\bar{o}$ (2005)]{kur05} Kurokawa, C. \& Kat$\bar{o}$, K. 2005, \prc, 71. 021301
\bibitem[Lucatello et al.(2005)]{luc05} Lucatello, S., Tsangarides, S., Beers, T. C., Carretta, E., Gratton, R. G., \& Ryan, S. G. 2005 \apj, 625, 825
\bibitem[Mitake et al.(1984)]{mit84} Mitake, S., Ichimaru, S., \& Itoh, N. 1984, \apj, 277, 375
\bibitem[Munakata et al.(1985)]{mun85} Munakata, H., Kohyama, Y., \& Itoh, N. 1985, \apj, 296, 197; erratum 
304, 580 (1986)
\bibitem[Nomoto et al.(1985)]{nom85} Nomoto, K., Thielemann, F. K., \& Miyaji, S. 1985, \aap, 149, 239
\bibitem[Picardi et al.(2004)]{pic04} Picardi, I., Chieffi, A., Limongi, M., Pisanti, O., Miele, G., Mangano, G., \& Imbriani, G. 2004 \apj, 609, 1035
\bibitem[Rogers \& Iglesias(1992)]{rog92}Rogers, F. J. \& Iglesias, C. A. 1992, \apjs, 79, 507
\bibitem[Rossi et al.(1998)]{ros98} Rossi, S., Beers, T.C.,
 \& Sneden, C. in ASP Conf. Ser. 165, The third Stromlo Symposium:
 The Galactic Halo, eds. B. K. Gibson, T. S. Axelrod, \& . E. Putman
 (San Francisco), 264. 
\bibitem[Schlattl et al.(2001)]{sch01} Schlattl, H., Cassisi, S., Salaris, M., \& Weiss, A. 2001, \apj, 559, 1082
\bibitem[Schlattl et al.(2002)]{sch02} Schlattl, H., Salaris, M., Cassisi, S., \& Weiss, A. 2002, \aap, 395, 77
\bibitem[Schwarzschild \& H\"arm(1965)]{sch65} Schwarzschild, M. \& H\"arm, R. 1965, \apj, 142, 855
\bibitem[Siess, Livio \& Lattanzio(2002)]{sie02} Siess, L., Livio, M., \& Lattanzio, J. 2002, \apj, 570, 329
\bibitem[Slattery et al.(1980)]{sla80} Slattery, W. L., Doolen, G. D., \& DeWitt, H. E. 1980, \pra, 14, 840
\bibitem[Slattery et al.(1982)]{sla82} Slattery, W. L., Doolen, G. D., \& DeWitt, H. E. 1982, \pra, 26, 2255
\bibitem[Straniero(1988)]{str88} Straniero, O. 1988, \aaps, 76, 157
\bibitem[Suda(2003)]{sud03}Suda, T. 2003, PhD. Thesis
\bibitem[Suda et al.(2004)]{sud04}Suda, T., Aikawa, M., Machida, M. N., Fujimoto, M. Y., \& Iben, I. Jr. 2004, \apj, 611, 476
\bibitem[Sugimoto \& Fujimoto(1978)]{sug78}Sugimoto, D., \& Fujimoto, M. Y. 1978, \pasj, 30, 467
\bibitem[Thielemann et al.(1987)]{thi87} Thielemann, F. K., Arnould, M., \& Truran, J. W. 1987, in Advances in Nuclear Astrophysics, ed. E. Vangioni-Flam, J. Audouze, M. Casse, J.-P. Chieze, \& J. Tran Thanh Van(Gif-sur-Yvette: Ed. Fronti\`eres), 525
\bibitem[Wagner(1974)]{wag74} Wagner, R. L., 1974, \apj, 191, 173
\bibitem[Weiss et al.(2000)]{wei00} Weiss, A., Cassisi, S., Schlattl, H., \& Salaris, M. 2000, \apj, 553, 413
\bibitem[Weiss et al.(2004)]{wei04} Weiss, A., Schlattl, H., Salaris, M., \& Cassisi, S. 2004, \aap, 422, 217
\end{thebibliography}
\end{document}